\documentclass[prd,10pt,aps,preprintnumbers,twocolumn,nofootinbib,superscriptaddress,floatfix,amssymb,amsmath]{revtex4}

\begin{document}

\title{Two-component liquid model for the quark-gluon plasma}\thanks{Talk given by V.I.~Zakharov at
16th International Seminar on High Energy Physics (QUARKS-2010), Kolomna, Russia, 6-12 June, 2010.}

\author{M. N. Chernodub}\thanks{On leave from ITEP, Moscow, Russia.}
\affiliation{Laboratoire de Math\'ematiques et Physique Th\'eorique, Universit\'e Fran\c{c}ois-Rabelais Tours,
F\'ed\'eration Denis Poisson - CNRS, Parc de Grandmont, 37200 Tours, France}
\affiliation{Department of Physics and Astronomy, University of Gent, Krijgslaan 281, S9, B-9000 Gent, Belgium}
\author{Henri Verschelde}
\affiliation{Department of Physics and Astronomy, University of Gent, Krijgslaan 281, S9, B-9000 Gent, Belgium}
\author{V.I. Zakharov}
\affiliation{Institute for Theoretical and Experimental Physics,
B.~Cheremushkinskaya 25, Moscow, 117218, Russia}

\begin{abstract}
We  consider a two-component-liquid model, a la Landau, for the quark-gluon plasma.
Qualitatively, the model fits well some crucial observations concerning the plasma
properties. Dynamically, the model assumes the existence of an effective  scalar field which
is condensed. The existence of such a condensate is supported by lattice data.
We indicate a possible crucial test of the model by lattice simulations.
\end{abstract}

\date{\today}

\maketitle

\section{\label{sec:intro}Introduction}

The discovery of the strongly interacting quark-gluon plasma at RHIC\footnote{For details,
discussions and references see, e.g., reviews \cite{teaney}.} made a great impact on the
landscape of theoretical papers devoted to quantum chromodynamics.
There emerged a new problem of explaining
the exotic properties of the plasma.
It is as  fundamental and interesting as the confinement problem
and in fact the two problems are to be considered
in conjunction with each other.
Moreover, there is renewed interest in relativistic hydrodynamics, superfluidity
and, more generally, in applying the  holographic methods
to condensed-matter systems \cite{herzog}.

In this paper we consider the possibility
that a variation of the famous two-component model
of superfluidity applies directly to the quark-gluon
plasma\footnote{ The basic idea is the same as in our report~\cite{chernodub}.
Here, we extend the arguments and address the issue of crucial tests of the model.}.
Let us remind the reader, very briefly, the model itself.
The main point is that a volume element of the liquid cannot be characterized
any longer  by a single 4-velocity $u^{\mu}$ with $ (u^{\mu})^2=-1$. Instead there are two substances,
or motions with (normal) density $\rho_n$ and superfluid density $\rho_s$
and with independent 4-velocities, $u^{\mu}$ and $v^{\mu}$, respectively.
The total density is the sum of the two components:
\begin{equation}
\label{eq:rho:tot}
\rho_{\mathrm{tot}}~=~\rho_n+\rho_s\,.
\end{equation}
The superfluid component is described in terms of a scalar field $\phi$.
Which in the non-relativistic limit becomes the phase of the condensate
wave function.

Formally, the current and energy-momentum tensor
are written as follows\footnote{We follow here the notations of Ref. \cite{yarom}.
References to earlier paper can be found here as well.}:
\begin{eqnarray}\label{scalar}
j^{\nu} & = & \rho_nu^{\nu} + \rho_sv^{\nu}\\ \nonumber
T^{\nu\sigma} & = & (\epsilon+P)u^{\nu}u^{\sigma}+P\eta^{\nu\sigma}+\mu\rho_sv^{\nu}v^{\sigma}
~~,\nonumber
\end{eqnarray}
where $\eta^{\nu\sigma}=(-1,1,1,1)$, $\epsilon$ is the energy density of the normal component,
$P$ is the pressure and $\mu$ is the chemical potential. Note also that for simplicity
we approximated the energy-momentum tensor by the case of an ideal liquid.
The equations of motion are then
\begin{equation} \label{eq:eom}
\partial_{\nu}T^{\nu\sigma}~=~0\,, \quad \partial_{\nu}j^{\nu}~=~0\,, \quad u^{\nu}v_{\nu}~=~-1\,.
\end{equation}
The most important point about the model is, of course, our assumption of two independent
motions taking place at the same point. We will comment on this later, in the context of field theory.

The outline of the paper is as follows. First, in Sec.~2 we discuss the
qualitative features of the plasma and argue that the two-component model
fits the data qualitatively. Next, in Sec.~3 we consider the issue of
scalar fields in Yang-Mills theories. The
existence of scalar fields with certain properties is dynamically a prerequisite
for the validity of the model. We conclude that the lattice data rather support
the  existence of (effective) scalar fields. In Sec.~4 we propose a crucial
test of the model through measuring the correlator of components of the
energy-momentum tensor.

\section{Qualitative features }

It might be useful (for the purpose of model building) to reduce the
plasma properties to three points, namely, equation of state, viscosity
and the role of quantum effects.

\vskip 2mm
{\bf A.} The existence of the plasma was conjectured long time ago.
Moreover the equation of state of the plasma has been known
also since long since it  was established via numerical
experiments within the lattice formulation of QCD, for references see, e.g.,  \cite{teaney}.
It turns out that the equation of state is close to that
of an ideal gas of quark and gluons:
\begin{equation}\label{gas}
[\epsilon(T)]_{\mathrm{plasma}}~\approx~ \big(1-\delta\big) \, [\epsilon (T)]_{\mathrm{ideal~gas}}\,,
\end{equation}
where the correction $\delta\approx 0.15$, $\epsilon(T)$ is the energy density as function of
temperature and $[\epsilon (T)]_{\mathrm{ideal~ gas}}$ is
the energy density for non-interacting quarks and gluons.

Thus, the equation of state indicates that the plasma is close to an ideal gas.

\vskip 2mm
{\bf B}. The observation (\ref{gas}) produces  the illusion of simplicity of the properties of the
plasma. However, analysis of the data obtained at RHIC led to the conclusion that
the plasma   possesses the lowest viscosity $\eta$ among all the substances known
so far:
\begin{equation}\label{liquid}
\Big({\eta\over s}\Big)_{\mathrm{plasma}}~\approx~{1\over 4\pi},~~~
\end{equation}
where $s$ is the entropy density (introduced to measure the viscosity in dimensionless units).
The value of $1/4\pi$ is somewhat symbolical.
The actual value of $\eta$ might be larger, say $\eta/s\sim 0.4$ \cite{teaney} or even lower,
see \cite{romatschke}. The value $\eta/s=1/4\pi$ represents the conjectured lower limit \cite{son1}.

Anyhow, the viscosity observed for the plasma is the lowest one among all
the known liquids \cite{teaney}. Thus, measurements of the viscosity indicate that the
plasma is close to an ideal liquid (which is defined as having $\eta=0$).
Note that for the ideal gas the viscosity tends to infinity,
\begin{equation}
\Big({\eta\over s}\Big)_{\mathrm{ideal~gas}}~\to~\infty ~.
\end{equation}
More precisely, this ratio is inverse proportional to the  coupling
constant squared $\eta/s~\sim~1/\alpha_s^2$.

\vskip 2mm
{\bf C.} As a kind of variation of  point {\bf B}, one argues \cite{son1}
that such a low value of viscosity implies that quantum effects are
crucial and that the liquid cannot be, rigorously speaking, treated classically.
Indeed, based on estimates common in kinetics one readily finds that
$$\eta~\sim~ \tau_{\mathrm{relaxation}} \epsilon~~,$$
while for the entropy density one can use $s\sim k_Bn$
where $k_B$ is the Boltzmann constant, $\tau$ is the relaxation time,
$\epsilon$  is the energy density
and $n$ is the density of particles.
The central point is that
from the uncertainty principle the product of energy of a particle, $\epsilon/n$
times its life time, $\tau$ cannot be smaller that the Planck constant. Thus:
\begin{equation}\label{quantum}
{\eta\over s}~\sim~{\tau_{\mathrm{relaxation}}\over \tau_{\mathrm{quantum}}}~,
\end{equation}
where the "quantum time" $\tau_{\mathrm{quantum}}~\sim~ h/k_BT$.
Then the observation (\ref{liquid}) implies quantum nature of
the quark--gluon plasma.

It is a challenge to theory to explain  all  three observations, (\ref{gas}),  (\ref{liquid}),
(\ref{quantum})
which are apparently pointing in opposite directions.
Indeed, one starts at point {\bf A} with the idea that the plasma is an ideal gas
and ends up at point {\bf C} with a kind of a proof that the plasma is
in fact a quantum liquid.

It is amusing that it is quite straightforward to suggest a model
which allows -- on a qualitative level -- to unify all the would-be contradictory features of
the plasma \cite{chernodub}. We have in mind the two-component model
of superfluidity a la Landau.

Indeed, what is ``special'' about the viscosity? How is it possible to have
an equation of state close to that of the ideal gas and still a nearly
vanishing viscosity? Let us imagine that we are dealing with a two-component substance.
One of the components occupies a larger phase space, $c_1$ and is responsible for
the equation of state. The other one has a smaller phase space, $c_2$ but very small viscosity.
Then the total viscosity can still be small since, at least naively, to evaluate the
total viscosity one adds
inverse powers of the partial viscosities:
\begin{equation}\label{parallel}
{1\over \eta_{\mathrm{tot}}}~=~{c_1\over \eta_1}~+~{c_2\over \eta_2}~,
\end{equation}
where $c_{1,2}$ are normalized by $c_1+c_2=1$.
Indeed, the meaning of the viscosity $\eta$ is similar to that of resistance
and if we have two independent motions then we would apply the rule\footnote{Equation (\ref{parallel})
can be found in, e.g., in  old books on classical solutions \cite{old}. In more modern terms,
the example of the superfluidity itself might serve
as the best illustration to (\ref{parallel}). Indeed, the superfluid fraction can be small
while the whole liquid is superfluid. On more detailed level, some care should be exercised
since one has distinguish between viscosity with respect to a capillary motion and
with respect to rotations, for a recent exposition see, e.g., \cite{koh}.}  (\ref{parallel}).

Thus, the two-component model accommodates naturally points {\bf A, B} above.
Assuming one of the components be superfluid explains, as a bonus, the point {\bf C} as well.

Another point is worth emphasizing. In the non-relativistic case
the superfluid component evaporates at finite temperature $T_c$. The
physics behind this is readily understood. Indeed, at $T=0$ the superfluid component is related to
the condensate of particles with momentum ${\bf p}=0$. At non-vanishing temperature
the particles are excited by temperature. Because of the conservation of the number of particles
in the non-relativistic case, the superfluid component disappears at finite temperature.

In the relativistic case, that is in the absence of conservation of  particles,
the theoretical constraints on the phase space occupied by the superfluid component
are weaker. Indeed, even at $T\to \infty$ the non-perturbative component
in case of Yang-Mills theories vanishes only logarithmically:
\begin{equation}
\lim_{T\to\infty}{c_2(T)}~\sim~g^6_s(T)~\sim~{1\over (\ln T)^3}~,
\end{equation}
where $g^2_s(T)$ is the coupling of the original 4d theory.

Finally, we come to discuss the point {\bf C} above. Well, it is quite clear that if we assume superfluidity
then the bounds like (\ref{quantum})
could be violated. Indeed, in the superfluid case we have a condensate and the
whole counting of degrees of freedom in terms of the density $n$ breaks down, generally speaking.

To summarize, the two-liquid model explains very naturally all the three points {\bf A}-{\bf C}
which superficially look  self-contradictory.

\section{Scalar condensate}

\subsection{General constraints}

Dynamically, the validity of a superfluidity scenario depends crucially  on the existence
of an (effective) scalar~$\phi$, see the basic equations (\ref{scalar}).
This degree of freedom is kept in the hydrodynamic approximation and
is to be light, therefore. Moreover, in field-theoretic language
the only way to ensure lightness of a scalar is to have
spontaneous symmetry breaking, described by
a condensate
\begin{equation}\label{condition}
\langle \phi \rangle_{\mathrm{ground~state}}~\neq~0~.
\end{equation}
The phase of this condensate corresponds then to a new light degree of freedom.

The condition (\ref{condition}) looks very restrictive and, in more detail, assumes
in fact a number of constraints:
\begin{itemize}
\item[{\it a.}] The field $\phi$ is a complex field.

\item[{\it b.}] Nevertheless the condensate (\ref{condition}) should not violate
conservation of any known quantum number, like charge.

\item[{\it c.}] In case of superfluidity, one is to think rather in terms of
 a   {\it three-dimensional} field $\varphi({\bf r}) \equiv \arg \phi({\bf r})$ while its time derivative in the rest frame of the
 normal part of the fluid is determined by the chemical potential~$\mu$:
\begin{equation}\label{nonrelativistic}
\partial_t\varphi~=~\mu\,.
\end{equation}
\end{itemize}

Generalizations of (\ref{nonrelativistic}) to the case of relativistic plasma
are mentioned in the Introduction
[the third relation in Eq.~(\ref{eq:eom})].
It is not clear which charge could be associated with the chemical potential $\mu$.

\subsection{Thermal scalar}

If we consider the conditions {\bf A}--{\bf C} above in an abstract form, they look very
difficult to satisfy. It is then even more amusing that a 3d field with similar
properties arises naturally \cite{kogan} within the string approach to the
deconfinement phase transition and is commonly called
thermal scalar, for a concise review and further insights see \cite{kruczenski}.
The reservation is that the thermal scalar refers to the temperatures below $T_c$
while our prime interest is $T>T_c$.

One considers temperatures $T$ below and close to the temperature of the
Hagedorn transition $T_H$ that in critical string dimension $d=26$ coincides with the critical temperature $T_c$.
Below we neglect the difference between $T_H$ and $T_c$.
In the string picture $\beta_H\equiv 1/T_H=1/\alpha'$
where $(2\pi \alpha'\equiv l_s^{-2}$ is the string tension. At $T=T_H$
the statistical sum over the states diverges. The main observation is that
at small $|T-T_H|$ the sum is dominated by the contribution of a single degree of freedom,
that is a scalar field with mass
\begin{equation}\label{mass}
m_{\beta}^2~\simeq~{\beta_H(\beta-\beta_H)\over 2\pi^2(\alpha')^2}~~,
\end{equation}
In other words, at $T=T_H$ the mass is becoming tachyonic.

In more detail, it is convenient to use the polymer approach
to field theory of a scalar particle (see, e.g. \cite{polyakov}) so that the action associated
with a trajectory of length $L$  is $S~=~M\cdot L$ where $M$ is the bare mass.
The trajectories are random walks with renormalized mass.
The free energy of the thermal scalar can be represented as a sum over random walks
and the final expression reduces to:
\begin{equation}\label{polymer}
F~=\beta\ln Z~=\beta\int\limits_0^{\infty}{dL\over L}{\exp (-m_{\beta}^2l_sL)\over (l_sL)^{d/2}}~~,
\end{equation}
where $d$ is the number of spatial coordinates, in our case $d=3$.
Expression (\ref{polymer}) is quite generic to the polymer approach.
A specific feature of (\ref{polymer})  is that $l_s$ plays the role of the length
of the links and is fixed in terms of the string tension.

The crucial point is that the free energy of the thermal scalar is exactly the partition
function for a single static string with tension $1/(2\pi\alpha')$. Moreover, the
single string dominates the free energy of a gas of strings.

\subsection{Scalar particles at $T>T_c$}

What happens to the thermal scalar at $T>T_c$ is an open question.
Consider first the case of a second order phase transition
which is relevant to the SU(2) gauge group.
Then we would expect that the thermal scalar is condensed at $T>T_c$.
Such a scenario is typical for the percolation picture,
which is a realization of the second-order-phase-transition scenario,
see, e.g. \cite {grimmelt}
The basic features can be understood from Eq. (\ref{polymer}).
At $m_{\beta}^2=0$ the exponential suppression of very
large lengths $L$ disappears. However, the integral over $L$ is still
divergent in the ultraviolet, not in the infrared. This means that
small clusters with $L\sim l_s$  dominate. The probability of having infinite length
is suppressed by a power of $L$ at $L\to \infty$. For a tachyonic mass
there emerges an infinite cluster. However, its density is suppressed as
a power of $m_{\beta}^2$ and small for temperatures above and close to $T_c$.
In field theoretic language appearance of the infinite cluster means
condensation of the field, $\langle \phi \rangle \neq 0$.

Imagine that the thermal scalar is indeed condensed at $T>T_c$.
Then, remarkably enough, the conditions we formulated above are
satisfied. Indeed,
\begin{itemize}
\item[a)] The thermal scalar is a complex field. It is encoded in the fact
that the integration in (\ref{polymer}) is over closed loops which
means a complex field in the polymer language.

\item[b)] The thermal scalar is associated with topological quantum number
which is a wrapping around the compactified time direction (due to finite
temperature).

\item[c)] The thermal scalar is a 3d scalar field, as it follows from
 the representation (\ref{polymer}).

\item[d)] Concerning the chemical potential $\mu$.
As is emphasized in \cite{kruczenski}, near the phase transition all string configurations
are time-oriented. In terms of random-walk formalism for a scalar particle
time orientation of the walk means chemical potential.
\end{itemize}

Nowadays, it is common to consider dual models of Yang-Mills theories
in terms of strings living in extra dimensions with non-trivial
geometry. The thermal scalar at temperatures below and close to $T_c$
is generic to such models as well,
see \cite{kruczenski} and references therein. One would not claim, however, that the most naive
version of the condensation of the thermal scalar is realized within
this scenario. Rather, the phase transition is a change of geometry
in the extra dimensions.

However, the scalar fields at $T>T_c$ are resurrected in another disguise.
Namely, one predicts existence of defects of various dimensions,
see in particular \cite{gorsky}. At $T>T_c$ the models predict existence
of time-oriented strings. Their 3d projection then looks as
 trajectories and correspond indeed to scalar 3d particles.
 There are independent  lattice data which seem to support
 the validity of this prediction \cite{nakamura}.

 To summarize, there is strong evidence that at $T{>}T_c$ there exists an effective
 3d scalar
 field condensed in the thermal vacuum of QCD. The existence of such a scalar is a
 necessary condition for the validity of the two-component model.

 \section{Possible crucial test of the model}

 The considerations given above demonstrate that the two-component model
 of the quark-gluon plasma does not contradict existing data. One cannot claim,
 however, that the model is indeed validated by the data.

 A crucial test of the model could performed through lattice measurements
 of a correlator of components of the energy-momentum tensor $T^{ti},i=1,2,3$,
 where the index $t$ stands for the Euclidean time direction.
 In more detail, consider the retarded Green's function defined as:
 \begin{equation}\label{corr}
 G_R^{tj,ti}(k)~\equiv~i\int d^4x\, e^{-ikx}\theta(t)\langle[T^{tj}(x),T^{ti}(0)]\rangle\,.
 \end{equation}
 Moreover, concentrate on the case of vanishing frequency, $k_0=0$.
 There are two independent form factors, corresponding to transverse and longitudinal waves,
 \begin{equation}\label{GR}
 G_R^{tj,ti}(0, {\bf k})~=~{k^ik^j\over {\bf k}^2}G^{L}_R({\bf k})+\Big(\delta^{ij}-
 {k^ik^j\over {\bf k}^2}\Big)G_R^T({\bf k}).
 \end{equation}

 Contribution of the superfluid component to the form factors $G_R^{L,T}$ has been discussed in many
 papers and textbooks. Here, we quote the result of the paper \cite{yarom}
 which includes also relativistic corrections:
 \begin{equation}\label{limit}
 \begin{array}{rlc}
 \lim\limits_{{\bf k}\to 0} G_R^T({\bf k}) & = & -(sT+\mu\rho_n), \\[3mm]
 \lim\limits_{{\bf k}\to 0} G_R^L({\bf k}) & = & -(sT+\mu\rho_{\mathrm{tot}}),
 \end{array}
 \end{equation}
where $s$ is the entropy density, $T$ is the temperature, $\mu$ is the chemical potential,
$\rho_\mathrm{tot}= \rho_n+\rho_s$ is the total density~(\ref{eq:rho:tot}), while $\rho_n$ and $\rho_s$ are the densities
of the normal and superfluid components, respectively.

Equations~(\ref{GR}) and (\ref{limit}) lead to the following result for zeroth
Matsubara frequency of the correlator~(\ref{corr})
\begin{equation}\label{test}
\lim_{{\bf k}\to 0}{G_R^{tj,ti}(0,{\bf k})}  =
- \delta^{ij} (s T + \mu \rho_n)
+ \mu \rho_s{k^ik^j\over {\bf k}^2}\,.
\end{equation}
It is only the superfluid component that leads to the non-analyticity in the limit of small spatial momenta~${\bf k}$.
And it is only the superfluid component that leads to appearance of the off-diagonal terms in the correlator~(\ref{test}).

Thus, we propose to test the possible presence of the superfluid component by evaluating the off-diagonal components
of the correlator~(\ref{test}).
Note that the proposed crucial test of the two-component model (\ref{test}) refers to static quantities,
corresponding to {\it exactly} zero temporal momentum in Minkowski space, $k_0=0$, and, consequently, to the zero
Matsubara frequency, $\omega_n = 0$,  on the lattice.
Since there is no time (or frequency) dependence, the continuation from the Euclidean
to Minkowski space is straightforward,
and no analytical continuation in the low--frequency region is required.
Thus, the prediction of the model, $\rho_s\neq 0$, can directly be tested on the lattice.

\section{Conclusions}

It is amusing that the known qualitative features of the quark-gluon plasma seem to favor a
two-component model of superfluidity for the plasma. In terms of field theory, the model
implies the condensation of an effective 3d scalar field. This consequence of the model seems
to be qualitatively supported by the lattice data as well.

A crucial test of the model could be performed through the search for the non-analyticity in the
spatial off-diagonal components of the correlator~(\ref{test}) of the energy-momentum tensor on the lattice.
This property can be tested directly in lattice simulations of Yang-Mills theories.

\acknowledgments
The work of MNC has been partially supported by the French Agence Nationale de la Recherche project ANR-09-JCJC ``HYPERMAG''.

\end{document}